\documentclass[12pt,showkeys,showpacs] {revtex4}
\usepackage{pifont}
\usepackage{amsmath}
\usepackage{graphicx}
\usepackage{cases}

\begin{document}

\title[Short Title]{Accelerating adiabatic quantum transfer for three-level $\Lambda$-type structure systems via picture transformation}

\author{Yi-Hao Kang$^{1,2}$}
\author{Qi-Cheng Wu$^{1,2}$}
\author{Ye-Hong Chen$^{1,2}$}
\author{Zhi-Cheng Shi$^{1,2}$}
\author{Jie Song$^{3}$}
\author{Yan Xia$^{1,2,}$\footnote{E-mail: xia-208@163.com}}

\affiliation{$^{1}$Department of Physics, Fuzhou University, Fuzhou 350116, China\\
             $^{2}$Fujian Key Laboratory of Quantum Information and Quantum Optics (Fuzhou University), Fuzhou 350116, China\\
             $^{3}$Department of Physics, Harbin Institute of Technology, Harbin 150001, China}

\begin{abstract}
In this paper, we investigate the quantum transfer for the system
with three-level $\Lambda$-type structure, and construct a shortcut
to the adiabatic passage via picture transformation to speed up the
evolution. We can design the pulses directly without any additional
couplings. Moreover, by choosing suitable control parameters, the
Rabi frequencies of pulses can be expressed by the linear
superpositions of Gaussian functions, which could be easily realized
in experiments. Compared with the previous works using the
stimulated Raman adiabatic passage, the quantum transfer can be
significantly accelerated with the present scheme.


\end{abstract}

\maketitle

\section{INTRODUCTION}

The three-level $\Lambda$-type system is known as a very important
model in quantum information processing (QIP). Many quantum
information tasks, such as the preparations of entanglement and the
operations of various quantum gates, can be implemented in physical
systems which are equivalent or approximately equivalent to
three-level systems with $\Lambda$-type structures
\cite{SongNJP6,ZhangSR5,yehongOC140,chenzhenSR6,lumeiPRA89,yehongPRA89,yehongSR5,xiaobinQIP14,wujiangQIP15,WeiQIP14,WuQIP,yehongPRA91,LonghiLPR3,LonghiJPB44,OrnigottiJPB41,RangelovPRA85}.
It is universally known that, to manipulate states of a three-level
quantum system with electromagnetic field, there exists two typical
methods, the $\pi$-pulse \cite{GolubevPRA90,ZhengPRL90} and the
adiabatic passage
\cite{ZhengPRL95,FewellAJP50,BergmannRMP70,VitanovARPC52,KralRMP79}.
These two methods hold their own advantages, but both reveal their
shortcomings. The $\pi$-pulse allows physical systems evolve
quickly, but the pulses should be controlled very accurately, which
bring challenges to experiments in some cases. On the other hand,
the adiabatic passage is famous for its robustness against the
imperfect operations and the deviations of the control parameters,
but badly limits the evolution speed of the systems, which makes the
systems more sensitive to some kinds of noise and decoherence
factors. For the sake of both high evolution speed and robustness, a
new technique named ``Shortcuts to adiabatic passage'' (STAP)
\cite{DemirplakJPCA107,DemirplakJCP129,TorronteguiAAMOP62,BerryJPA42,ChenPRL105,CampoPRL111,ChenPRA83,MugaJPB42,ChenPRL104,CampoSR2}
has been proposed.

The STAP suggests the system evolving in a controllable nonadiabatic
way, so that the adiabatic condition, which limits the evolution
speed of the system, can be abandoned. Besides, when the boundary
condition of the control parameters is well designed, the STAP is
also robust against the imperfect operations and the deviations of
the control parameters. Since the STAP combines the advantages of
both the $\pi$-pulse and the adiabatic passage, it has attracted
many interests of researches in different fields
\cite{TorronteguiPRA83,MugaJPB43,TorronteguiPRA85,MasudaPRA84,ChenPRA82,SchaffNJP13,ChenPRA84,TorronteguiNJP14,CampoPRA84,CampoEPL96,RuschhauptNJP14,SchaffPRA82,SchaffEPL93,ChenPRA86,SantosSR5,SantosPRA93,HenPRA91,SarandyQIP3,Coulamyarxiv,RamsarXiv,DeffnerPRX4,DuNC7,AnNC7}.
For example, Torrontegui \emph{et al.} \cite{TorronteguiNJP14} have
used STAP to transport Bose-Einstein condensates. Ruschhaupt
\emph{et al.} \cite{RuschhauptNJP14} have achieved a population
inversion in a two-level quantum system with STAP. Among these
schemes
\cite{DemirplakJPCA107,DemirplakJCP129,TorronteguiAAMOP62,BerryJPA42,ChenPRL105,CampoPRL111,ChenPRA83,MugaJPB42,ChenPRL104,CampoSR2,TorronteguiPRA83,MugaJPB43,TorronteguiPRA85,MasudaPRA84,ChenPRA82,SchaffNJP13,ChenPRA84,TorronteguiNJP14,CampoPRA84,CampoEPL96,RuschhauptNJP14,SchaffPRA82,SchaffEPL93,ChenPRA86,SantosSR5,SantosPRA93,HenPRA91,SarandyQIP3,Coulamyarxiv,RamsarXiv,DeffnerPRX4,DuNC7,AnNC7},
the method named ``transitionless quantum driving'' (TQD) (also
known as the ``counterdiabatic driving'')
\cite{DemirplakJPCA107,DemirplakJCP129,BerryJPA42,DuNC7,AnNC7} is
one of the famous methods for constructing STAP, whose idea is to
cancel the nonadiabatic transitions between the eigenstates for the
original Hamiltonian of the system by adding ``counterdiabatic''
(CD) terms. The CD terms can be calculated easily, and their
mathematic expressions are usually not too complex. For example,
Demirplak \emph{et al.} \cite{DemirplakJPCA107} have first used
counterdiabatic fields to accelerate adiabatic passages, and shown
that a population transfer between molecular states could be
perfectly achieved, which is a pioneering work of STAP. Moreover, Du
\emph{et al.} \cite{DuNC7} have experimentally shown that TQD could
be used to design pulses to construct STAP for cold atoms.
Furthermore, TQD has also be exploited by An \emph{et al.}
\cite{AnNC7} to experimentally realize trapped-ion displacement in
phase space. However, the CD terms sometimes play the roles as the
additional couplings which are hard to be realized in real
experiments. For example, it is indicated in many previous schemes
\cite{ChenPRL105,GiannelliPRA89,MasudaJPCA119,BasonNP8} that, for a
three-level $\Lambda$-type atom, the CD terms are the special
one-photon 1-3 pulse (the microwave field), which bring troubles to
the experimental realization.

To overcome the difficulties of TQD, many interesting schemes
\cite{BaksicPRL116,GaraotPRA89,OpatrnyNJP16,SaberiPRA90,TorronteguiPRA89,TorosovPRA87,TorosovPRA89,yehongPRA93,yehongarxiv,KangSR6,IbanezPRA87,IbanezPRL109,SongPRA93,HuangLPL13}
have been put forward. For example, Ib\'{a}\~{n}ez \emph{et al.}
\cite{IbanezPRL109,IbanezPRA87} have pointed out that a sequence of
shortcuts to adiabaticity can be built with similar way of TQD via
iterative interaction pictures. Subsequently, this method was used
in a three-level system with $\Lambda$-type structure by Song
\emph{et al.} \cite{SongPRA93}. They have shown that the
difficulties of TQD can be overcome, and the STAP can be constructed
by adjusting the Rabi frequencies of pulses in original Hamiltonian,
so the additional couplings are unnecessary. Chen \emph{et al.}
\cite{yehongarxiv} have also come up with an interesting idea to
construct an experimentally feasible Hamiltonian for a three-level
system by using multi-mode driving of a set of moving states. Baksic
\emph{et al.} \cite{BaksicPRL116} have proposed an interesting
scheme to speed up the quantum transfer for a three-level system
with a serial of dressed states. They have shown that canceling the
transitions between the chosen dressed states instead of the
transitions between the eigenstates of the original Hamiltonian can
avoid the difficulties of TQD, and the extra couplings are also
unnecessary.

Inspired by the works
\cite{BaksicPRL116,GaraotPRA89,OpatrnyNJP16,SaberiPRA90,TorronteguiPRA89,TorosovPRA87,TorosovPRA89,yehongPRA93,yehongarxiv,KangSR6,IbanezPRA87,IbanezPRL109,SongPRA93,HuangLPL13},
we propose an alternative scheme to accelerate the quantum transfer
for the system with three-level $\Lambda$-type structure. Different
from previous schemes, we directly investigate the dynamics of the
three-level $\Lambda$-type system and the solution of the
Schr\"{o}dinger equation with only one picture transformation. The
relationships of several parameters are studied. By designing these
parameters suitably, the Rabi frequencies of pulses can be directly
given, and they can be expressed by the linear superpositions of
Gaussian functions, which are feasible in experiments. Meanwhile,
the additional couplings are not required. In the end of the paper,
we perform the numerical simulations, which show that the present
scheme is effective. What is more, the quantum transfer can be
significantly accelerated by applying the scheme instead of that
with the stimulated Raman adiabatic passage.

\section{Accelerating adiabatic quantum transfer for three-level $\Lambda$-type structure system via picture transformation}

In this section, we start to introduce the method of the present
scheme. For a system with the three-level $\Lambda$-type structure,
the Hamiltonian has the general form as
\begin{eqnarray}\label{e1}
H_0(t)=\Omega_1(t)(|1\rangle\langle2|+|2\rangle\langle1|)+\Omega_2(t)(|3\rangle\langle2|+|2\rangle\langle3|),
\end{eqnarray}
where pulse with Rabi frequency $\Omega_1(t)$ ($\Omega_2(t)$) drives
the transition $|1\rangle\leftrightarrow|2\rangle$
($|2\rangle\leftrightarrow|3\rangle$). We suppose
\begin{eqnarray}\label{e2}
G_1=\left[%
\begin{array}{ccc}
  0 &\  1 &\  0 \\
  1 &\  0 &\  0 \\
  0 &\  0 &\  0 \\
\end{array}%
\right],\ \
G_2=\left[%
\begin{array}{ccc}
  0 &\  0 &\  0 \\
  0 &\  0 &\  1 \\
  0 &\  1 &\  0 \\
\end{array}%
\right],\ \
G_3=\left[%
\begin{array}{ccc}
  0 &\  0 &\  -i \\
  0 &\  0 &\  0 \\
  i &\  0 &\  0 \\
\end{array}%
\right],\ \
\end{eqnarray}
which satisfy the commutation relations $[G_1,G_2]=iG_3$,
$[G_2,G_3]=iG_1$ and $[G_3,G_1]=iG_2$. Assuming
$\Omega(t)=\sqrt{\Omega_1^2(t)+\Omega_2^2(t)}$,
$\tan\theta(t)=\Omega_1(t)/\Omega_2(t)$, the Hamiltonian in
Eq.~(\ref{e1}) can be rewritten as
\begin{eqnarray}\label{e3}
H_0(t)=\Omega(t)(\sin\theta G_1+\cos\theta G_2).
\end{eqnarray}

As the system possesses $SU(2)$ symmetry, we perform a picture
transformation as $|\psi_1(t)\rangle=B^{\dag}(t)|\psi_0(t)\rangle$
with the unitary operator $B(t)=e^{-i\epsilon(t)G_3}$, where
$|\psi_0(t)\rangle$ is the wave function in the original picture,
$|\psi_1(t)\rangle$ is the wave function after the picture
transformation, and $\epsilon(t)$ is a real parameter. With the
picture transformation, $H_0(t)$ will be transformed into
\begin{eqnarray}\label{e4}
H_1(t)&&=B^{\dag}(t)H_0(t)B(t)-iB^{\dag}(t)\dot{B}(t)\cr\cr&&
=\Omega(\sin\theta e^{i\epsilon G_3}G_1e^{-i\epsilon G_3}+\cos\theta
e^{i\epsilon G_3}G_2e^{-i\epsilon G_3})-\dot{\epsilon}G_3\cr\cr&&
=\Omega\sin(\theta+\epsilon)G_1+\Omega\cos(\theta+\epsilon)G_2-\dot{\epsilon}G_3.
\end{eqnarray}
In the following, we will prove that the Hamiltonian $H_1(t)$ in Eq.
(\ref{e4}) can be generated by the evolution operator in form of
$U_1(t)=e^{i\mu(\sin\varphi G_1+\cos\varphi G_2)}$ with parameters
$\mu(t)$ and $\varphi(t)$.

At the beginning, we assume $M(t)=(\sin\varphi G_1+\cos\varphi
G_2)$. The operator $M$ has three eigenstates
\begin{eqnarray}\label{e5}
&&|\xi_0\rangle=\cos\varphi|1\rangle-\sin\varphi|3\rangle,\cr\cr&&
|\xi_+\rangle=\frac{1}{\sqrt{2}}(\sin\varphi|1\rangle+|2\rangle+\cos\varphi|3\rangle),\cr\cr&&
|\xi_-\rangle=\frac{1}{\sqrt{2}}(\sin\varphi|1\rangle-|2\rangle+\cos\varphi|3\rangle),
\end{eqnarray}
corresponding to the eigenvalues 0, 1 and -1, respectively. It is
obviously that
\begin{eqnarray}\label{e6}
&&M(t)=|\xi_+\rangle\langle\xi_+|-|\xi_-\rangle\langle\xi_-|,\cr\cr&&
M^n(t)=|\xi_+\rangle\langle\xi_+|+(-1)^n|\xi_-\rangle\langle\xi_-|,\cr\cr&&
U_1(t)=e^{i\mu
M}=\sum\limits_{n=0}^{\infty}\frac{i^n\mu^nM^n}{n!}=|\xi_0\rangle\langle\xi_0|+e^{i\mu}|\xi_+\rangle\langle\xi_+|+e^{-i\mu}|\xi_-\rangle\langle\xi_-|,
\end{eqnarray}
and
\begin{eqnarray}\label{e7}
&&|\dot{\xi}_0\rangle=-\frac{\dot{\varphi}}{\sqrt{2}}(|\xi_+\rangle+|\xi_-\rangle),\cr\cr&&
|\dot{\xi}_+\rangle=|\dot{\xi}_-\rangle=\frac{\dot{\varphi}}{\sqrt{2}}|\xi_0\rangle.
\end{eqnarray}
Therefore, we can further obtain
\begin{eqnarray}\label{e8}
i\dot{U_1}(t)U_1^{\dag}(t)&&=-(\dot{\mu}\sin\varphi+\dot{\varphi}\sin\mu\cos\varphi)G_1
+(\dot{\varphi}\sin\mu\sin\varphi-\dot{\mu}\cos\varphi)G_2
-\dot{\varphi}(1-\cos\mu)G_3\cr\cr&&
=\gamma\sin(\varphi-\delta-\pi/2)G_1+\gamma\cos(\varphi-\delta-\pi/2)G_2-\dot{\varphi}(1-\cos\mu)G_3,
\end{eqnarray}
where, $\gamma=\sqrt{\dot{\mu}^2+\dot{\varphi}^2\sin^2\mu}$ and
$\tan\delta=\dot{\mu}/(\dot{\varphi}\sin\mu)$. Comparing
Eq.~(\ref{e8}) with Eq.~(\ref{e4}), we have
\begin{eqnarray}\label{e9}
&&\Omega=\gamma=\sqrt{\dot{\mu}^2+\dot{\varphi}^2\sin^2\mu},\cr\cr&&
\theta+\epsilon=\varphi-\delta-\pi/2,\cr\cr&&
\dot{\epsilon}=\dot{\varphi}(1-\cos\mu).
\end{eqnarray}

On the other hand, we assume the initial time is $t_i=0$ and the
final time is $t_f=T$. If  $|\psi_0(0)\rangle=|1\rangle$,
$\epsilon(0)=\varphi(0)=0$, we have
$|\psi_1(0)\rangle=|\psi_0(0)\rangle=|1\rangle$. With the evolution
operator $U_1(t)$, we obtain the wave function in the transformed
picture as
\begin{eqnarray}\label{e10}
|\psi_1(t)\rangle&&=U_1(t)|\psi_1(0)\rangle\cr\cr&&
=(|\xi_0\rangle\langle\xi_0|+e^{i\mu}|\xi_+\rangle\langle\xi_+|+e^{-i\mu}|\xi_-\rangle\langle\xi_-|)[\cos\varphi|\xi_0\rangle+\frac{1}{\sqrt{2}}\sin\varphi(|\xi_+\rangle+|\xi_-\rangle)]\cr\cr&&
=\cos\varphi|\xi_0\rangle+\frac{1}{\sqrt{2}}\sin\varphi(e^{i\mu}|\xi_+\rangle+e^{-i\mu}|\xi_-\rangle)\cr\cr&&
=(\cos^2\varphi+\sin^2\varphi\cos\mu)|1\rangle+i\sin\varphi\sin\mu|2\rangle+\sin\varphi\cos\varphi(\cos\mu-1)|3\rangle.
\end{eqnarray}
Moving back to the original picture, the wave function is
\begin{eqnarray}\label{e11}
|\psi_0(t)\rangle&&=B(t)|\psi_1(t)\rangle\cr\cr&&
=[\cos\epsilon(\cos^2\varphi+\sin^2\varphi\cos\mu)-\sin\epsilon\sin\varphi\cos\varphi(\cos\mu-1)]|1\rangle+i\sin\varphi\sin\mu|2\rangle+\cr\cr&&
[\cos\epsilon\sin\varphi\cos\varphi(\cos\mu-1)+\sin\epsilon(\cos^2\varphi+\sin^2\varphi\cos\mu)]|3\rangle.
\end{eqnarray}
With Eq.~(\ref{e9}) and Eq.~(\ref{e11}), we can design the control
parameters $\varphi$, $\epsilon$ and $\mu$ to realize a quantum
transfer with experimental feasible pulses.

As an example, we design a set of parameters and perform numerical
simulations to show the effectiveness of the present scheme. For
simplicity, we assume $\dot{\mu}\equiv0$, so that $\kappa=1-\cos\mu$
is a constant ($0\leq\kappa\leq2$). Then we have
$\epsilon=\kappa\varphi$, $\Omega=|\dot{\varphi}\sin\mu|$,
$\delta=0$, $\theta=(1-\kappa)\varphi-\pi/2$ and the wave function
in Eq.~(\ref{e11}) will become
\begin{eqnarray}\label{e12}
|\psi_0(t)\rangle=&&[\cos(\kappa\varphi)(1-\kappa\sin^2\varphi)+\kappa\sin(\kappa\varphi)\sin\varphi\cos\varphi]|1\rangle+i\sqrt{2\kappa-\kappa^2}\sin\varphi|2\rangle+\cr\cr&&
[\sin(\kappa\varphi)(1-\kappa\sin^2\varphi)-\kappa\cos(\kappa\varphi)\sin\varphi\cos\varphi]|3\rangle.
\end{eqnarray}
Assuming that we desire a quantum transfer
$|1\rangle\rightarrow|3\rangle$, we should have $2\kappa-\kappa^2=0$
or $\varphi(T)=m\pi$ ($m=0,\pm1,\pm2,\cdots$). It is obviously that
when $2\kappa-\kappa^2=0$, which gives $\kappa=0$ or $\kappa=2$, the
result is $|\psi_0(t)\rangle\equiv|1\rangle$. That means the quantum
transfer can not be realized when $2\kappa-\kappa^2=0$. Therefore,
we select $\varphi(T)=\pi$ here. So we have
\begin{eqnarray}\label{e13}
|\psi_0(T)\rangle=\cos(\kappa\pi)|1\rangle+\sin(\kappa\pi)|3\rangle.
\end{eqnarray}
In order to realize the quantum transfer
$|1\rangle\rightarrow|3\rangle$, we choose $\kappa=1/2$, i.e.,
$\mu=\pi/3$. Then the following results can be obtained:
$\Omega(t)=\frac{\sqrt{3}}{2}|\dot{\varphi}|$,
$\theta(t)=\frac{\varphi-\pi}{2}$. For the sake of robustness
against deviation of operation time, the boundary condition
$\Omega(0)=\Omega(T)=0$ is advisable. Therefore, we choose
\begin{eqnarray}\label{e14}
&&\varphi(t)=\frac{\pi}{2}[1-\cos(\frac{\pi t}{T})],\ \ \ \
\dot{\varphi}(t)=\frac{\pi^2}{2T}\sin(\frac{\pi t}{T}).
\end{eqnarray}

Until now, the only question remained is that the Rabi frequencies
$\Omega_1(t)=\Omega(t)\sin\theta(t)$ and
$\Omega_2(t)=\Omega(t)\cos\theta(t)$ are still too complex for the
experimental realization. For the sake of the experimental
feasibility, we apply the curve fitting to $\Omega_1(t)$ and
$\Omega_2(t)$, and obtain two replacing Rabi frequencies
$\widetilde{\Omega}_1(t)$ and $\widetilde{\Omega}_2(t)$ as
\begin{eqnarray}\label{e15}
&&\widetilde{\Omega}_1(t)=\zeta_{11}e^{-[(t-\tau_{11})/\chi_{11}]^2}+\zeta_{12}e^{-[(t-\tau_{12})/\chi_{12}]^2},\cr\cr&&
\widetilde{\Omega}_2(t)=\zeta_{21}e^{-[(t-\tau_{21})/\chi_{21}]^2}+\zeta_{22}e^{-[(t-\tau_{22})/\chi_{22}]^2},
\end{eqnarray}
for $\Omega_1(t)$ and $\Omega_2(t)$, respectively, where,
\begin{eqnarray}\label{e16}
&&\zeta_{11}=-3.194/T,\ \zeta_{12}=-1.275/T,\ \zeta_{21}=3.194/T,\
\zeta_{22}=1.275/T,\cr\cr&& \tau_{11}=0.4396T,\ \tau_{12}=0.2159T,\
\tau_{21}=0.5604T,\ \tau_{22}=0.7841T,\cr\cr&& \chi_{11}=0.2476T,\
\chi_{12}=0.1581T,\ \chi_{21}=0.2476T,\ \chi_{22}=0.1581T.
\end{eqnarray}
Here, $\zeta_{\alpha\beta}$ ($\alpha,\beta=1,2$) is the pulse
amplitude of the $\beta$th component in pulse $\Omega_\alpha(t)$,
$\tau_{\alpha\beta}$ describes the extreme point of the $\beta$th
component in pulse $\Omega_\alpha(t)$, and $\chi_{\alpha\beta}$
controls the width of the $\beta$th component in pulse
$\Omega_\alpha(t)$. To compare $\Omega_1(t)$ ($\Omega_2(t)$) and
$\widetilde{\Omega}_1(t)$ ($\widetilde{\Omega}_2(t)$), we plot
$|\Omega_1(t)|$ ($\Omega_2(t)$) and $|\widetilde{\Omega}_1(t)|$
($\widetilde{\Omega}_2(t)$) versus $t/T$ in Fig. 1 (a) (Fig. 1 (b)).
Seen from Fig. 1, $\Omega_1(t)$ ($\Omega_2(t)$) and
$\widetilde{\Omega}_1(t)$ ($\widetilde{\Omega}_2(t)$) are very close
to each other. Besides, the pulse amplitude
$\widetilde{\Omega}_0=\max\limits_{0\leq t\leq
T}\{\widetilde{\Omega}_1(t),\widetilde{\Omega}_2(t)\}$ is only about
$3.5/T$.
\begin{figure}
\scalebox{0.6}{\includegraphics[scale=1]{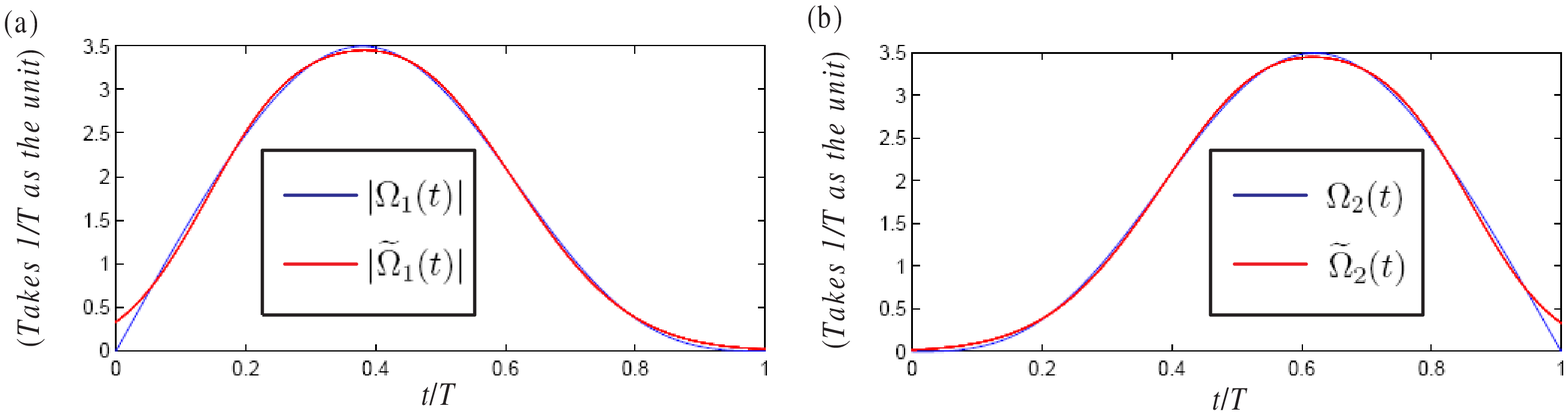}} \caption{(a)
Comparison between $|\Omega_1(t)|$ and $|\widetilde{\Omega}_1(t)|$
(versus $t/T$). (b) Comparison between $\Omega_2(t)$ and
$\widetilde{\Omega}_2(t)$ (versus $t/T$).}
\end{figure}
Moreover, the population $P_j(t)=|\langle j|\psi_0(t)\rangle|^2$ of
state $|j\rangle$ ($j=1,2,3$) is plotted in Fig. 2.
\begin{figure}
\scalebox{0.8}{\includegraphics[scale=1]{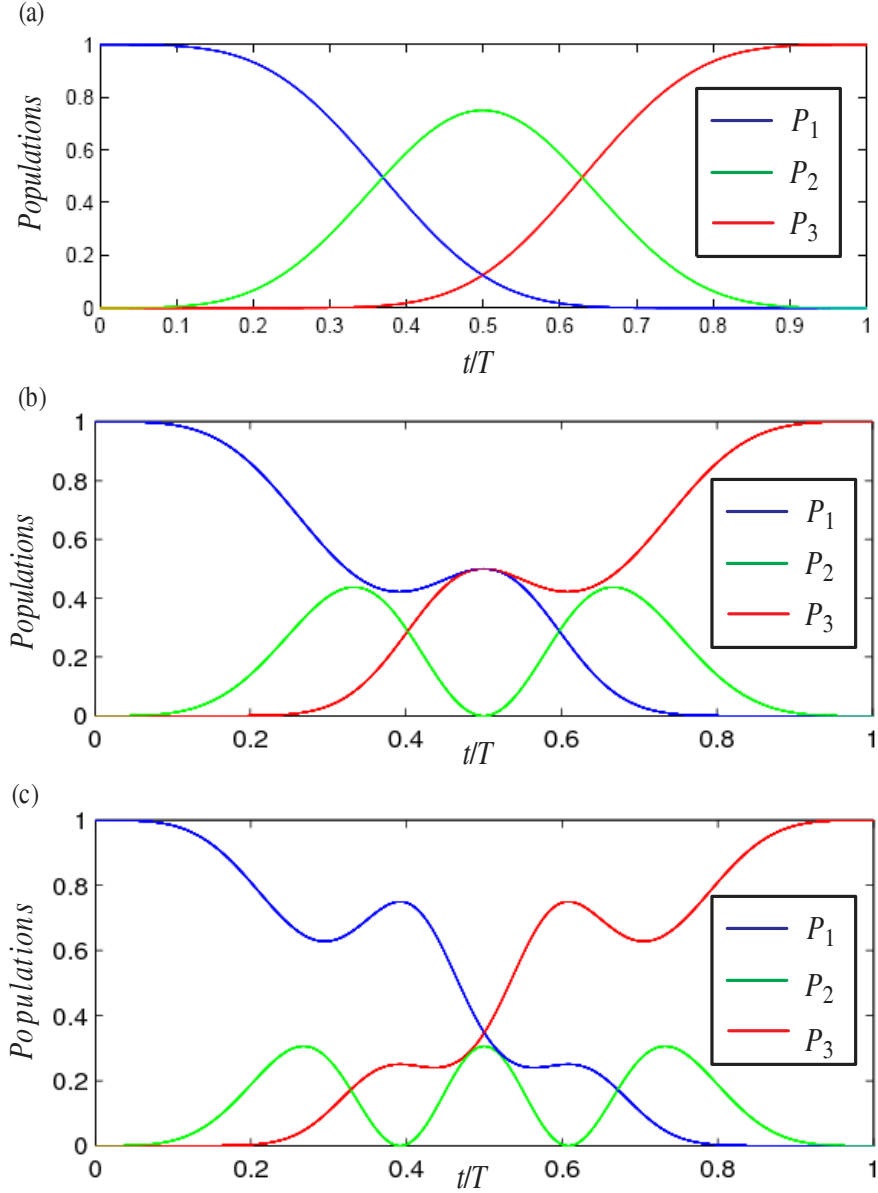}} \caption{The
population $P_j$ state of $|j\rangle$ versus $t/T$: (a)
$\varphi(T)=\pi$, $\kappa=1/2$; (b) $\varphi(T)=2\pi$, $\kappa=1/4$;
(c) $\varphi(T)=3\pi$, $\kappa=1/6$.}
\end{figure}
As shown in Fig. 2(a), $P_3$ increases from 0 to 1 during the
evolution, so the quantum transfer $|1\rangle\rightarrow|3\rangle$
can be achieved successfully. This proves that the method of the
scheme and the replacing pulses in Eq.~(\ref{e15}) are both
effective. Fig. 2(a) also shows that $P_2$, the population of the
intermediate state $|2\rangle$, reaches its maximal value
$P_{2max}=2\kappa-\kappa^2=0.75$ at $t=T/2$. If we want to decrease
$P_2$, we can increase $|\varphi(T)|$, so that a smaller $\kappa$
can be chosen. For example, if we choose $\varphi(T)=2\pi$, then
$\kappa=1/4$ can be adopted, so the maximal value of $P_2$ is
$P_{2max}=2\kappa-\kappa^2=0.4375$ (See Fig. 2(b)); if we choose
$\varphi(T)=3\pi$, then $\kappa=1/6$ is available, so the maximal
value of $P_2$ is $P_{2max}=2\kappa-\kappa^2=0.3056$ (See Fig.
2(c)). However, increasing $|\varphi(T)|$ requires us to increase
the maximal value of $|\dot{\varphi}|$, since $\max\limits_{0\leq
t\leq T}|\dot{\varphi}(t)|\geq|\varphi(T)|/T$ (See Table I). As a
result, $\widetilde{\Omega}_0$ should also be increased. To have a
relative high evolution speed, the product
$\widetilde{\Omega}_0\times T$ (of the pulse amplitude
$\widetilde{\Omega}_0$ and the total interaction time $T$) is the
smaller the better. Because when $\widetilde{\Omega}_0$ is fixed
(e.g. $\widetilde{\Omega}_0$ reaches the upper limit of the system),
a smaller product $\widetilde{\Omega}_0\times T$ means a short
interaction time $T$. On the other hand, in some cases, $P_2$ is
required to be restrained in order to decrease the dissipation.
Therefore, in real systems, one should choose a suitable value of
control parameters for higher evolution speed and less dissipation.
\begin{center}{\bf
Table I. The pulse amplitude $\widetilde{\Omega}_0$ and the maximal
value of intermediate state's population $P_{2max}$ with
corresponding $|\varphi(T)|$.\ \ \ \ \ \ \ \ \ \ \ \ \ \ \ \ \ \ \ \
\ \ \ }{\small
\begin{tabular}{ccc}\hline\hline
\ \ \ \ \ \ \ $|\varphi(T)|$ \ \ \ \ \ \ \ \ \ \ \ \ \ \ \ \ \ \ \ \ \ \ \ &\ \ \ \ \ $\widetilde{\Omega}_0$ \ \ \ \ \ \ \ \ \ \ \ \ \ \ \ \ \ \ \ \ \ &$P_{2max}$\\
\hline $\ \ \ \ \ \ \pi\ \ \ \ \ \ \ \ \ \ \ \ \ \ \ \ \ \ \ \ \ \ \
\ $&$ 3.5/T\
\ \ \ \ \ \ \ \ \ \ \ \ \ \ \ $&$0.75$\\
$2\pi\ \ \ \ \ \ \ \ \ \ \ \ \ \ \ \ \ \ \ $&$ 6.2/T\ \ \ \ \ \
\ \ \ \ \ \ \ \ \ \ $&$0.4375$\\
$3\pi\ \ \ \ \ \ \ \ \ \ \ \ \ \ \ \ \ \ \ $&$ 8.0/T\ \ \ \ \ \
\ \ \ \ \ \ \ \ \ \ $&$0.3056$\\
$4\pi\ \ \ \ \ \ \ \ \ \ \ \ \ \ \ \ \ \ \ $&$ 9.5/T\ \ \ \ \ \ \
\ \ \ \ \ \ \ \ \ $&$0.2344$\\
$5\pi\ \ \ \ \ \ \ \ \ \ \ \ \ \ \ \ \ \ \ $&$ 10.7/T\ \ \ \ \
\ \ \ \ \ \ \ \ \ \ \ $&$0.1900$\\
$6\pi\ \ \ \ \ \ \ \ \ \ \ \ \ \ \ \ \ \ \ $&$ 11.8/T\ \ \ \ \
\ \ \ \ \ \ \ \ \ \ \ $&$0.1597$\\
$7\pi\ \ \ \ \ \ \ \ \ \ \ \ \ \ \ \ \ \ \ $&$ 12.8/T\ \ \ \
\ \ \ \ \ \ \ \ \ \ \ \ $&$0.1378$\\
\hline \hline
\end{tabular} }
\end{center}

Now, we would like to show that the quantum transfer can be
significantly accelerated by using the present scheme. As a
comparison, the stimulated Raman adiabatic passage (STIRAP) is also
exploited to implement the quantum transfer. According to STIRAP
method, the system evolves through the dark state
$|\Psi_{dark}(t)\rangle=\frac{1}{\sqrt{\Omega_{1}^2(t)+\Omega_{2}^2(t)}}(\Omega_{2}(t)|1\rangle-\Omega_{1}(t)|3\rangle)$
of Hamiltonian $H_0(t)$ shown in Eq.~(\ref{e1}). By setting boundary
condition
\begin{eqnarray}\label{e17}
\lim\limits_{t\rightarrow-\infty}=\frac{\Omega_{1}(t)}{\Omega_{2}(t)}=0,\
\lim\limits_{t\rightarrow+\infty}=\frac{\Omega_{2}(t)}{\Omega_{1}(t)}=0,
\end{eqnarray}
one can design the Rabi frequencies $\Omega_{1}$ and $\Omega_{2}$ as
following
\begin{eqnarray}\label{e18}
&&\Omega_{1}(t)=\Omega_0\exp[-(\frac{t-t_0-T/2}{t_c})^2],\cr\cr&&
\Omega_{2}(t)=\Omega_0\exp[-(\frac{t+t_0-T/2}{t_c})^2],
\end{eqnarray}
where $\Omega_0$ denotes the pulse amplitude for STIRAP, $t_0$ and
$t_c$ are two related parameters. Setting $t_0=0.15T$ and
$t_c=0.2T$, Rabi frequencies $\Omega_{1}$ and $\Omega_{2}$ can well
satisfy the boundary condition in Eq.~(\ref{e17}). We plot
$1-P_3(T)$ versus $\Omega_0$ for the STIRAP method in Fig. 3. As
shown in Fig. 3, with STIRAP, for obtaining an enough high
population of state $|3\rangle$, one should have $\Omega_0\geq45/T$.
When $\Omega_0=70/T$, $1-P_3(T)$ is 0.0002. For the present scheme,
we have $1-P_3(T)\leq3.714\times10^{-5}$ with
$\widetilde{\Omega}_0=3.5/T$. But for STIRAP, when $\Omega_0=3.5/T$,
$1-P_3(T)=0.9906$ due to the great violation of the adiabatic
condition. As we mentioned above in this section, for a relatively
high evolution speed, the product of the pulse amplitude and the
total interaction time is the smaller the better. Therefore, using
the present scheme, the quantum transfer can be significantly
accelerated.
\begin{figure}
\scalebox{0.8}{\includegraphics[scale=1]{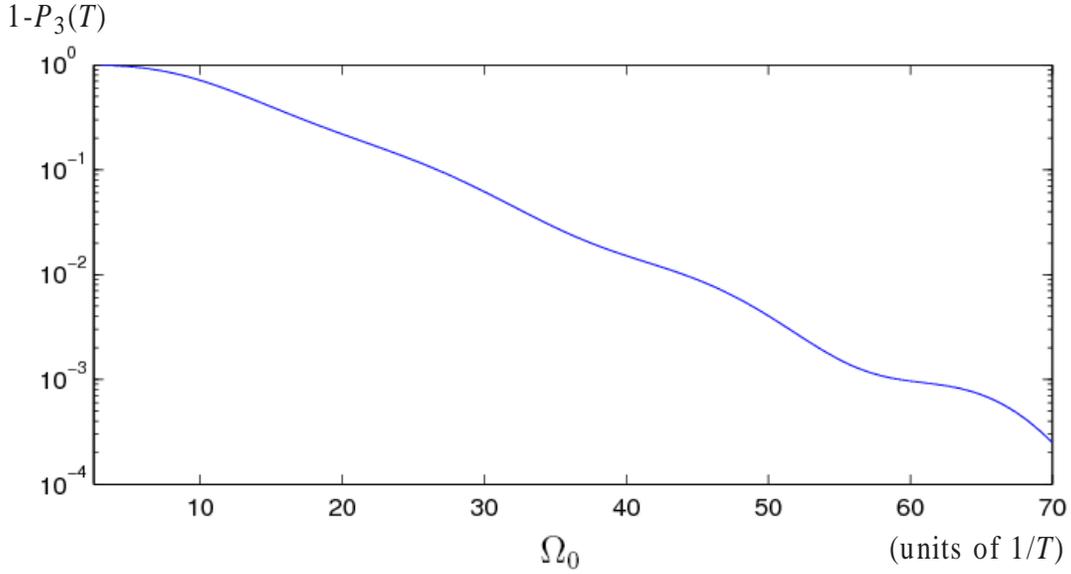}}
\caption{$1-P_3(T)$ versus $\Omega_0$ for the STIRAP method.}
\end{figure}

At the end of this section, let us check the robustness of the
scheme with some numerical simulations. Firstly, we would like to
show the robustness of the scheme against the parameters' errors
caused by the imperfect operations. Here, we consider the errors
$\delta T$, $\delta\widetilde{\Omega}_1$ and
$\delta\widetilde{\Omega}_2$ of the total interaction time $T$, the
Rabi frequencies of pulses $\widetilde{\Omega}_1$ and
$\widetilde{\Omega}_2$, respectively. Before we perform the
numerical simulations, we assume that $T'=T+\delta T$ is the
erroneous total interaction time. $P_3(T')$ versus
$\delta\widetilde{\Omega}_1/\widetilde{\Omega}_1$ and
$\delta\widetilde{\Omega}_2/\widetilde{\Omega}_2$ are shown by the
blue crosses and the solid-red line in Fig. 4, respectively. And
$P_3(T')$ versus $\delta T/T$ is plotted by the dashed-green line in
Fig. 4.
\begin{figure}
\scalebox{0.8}{\includegraphics[scale=0.8]{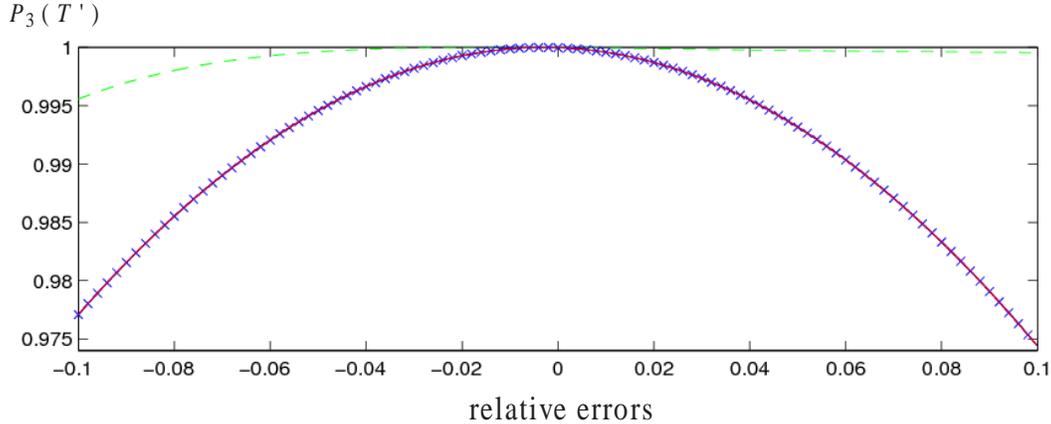}}
\caption{$P_3(T')$ versus relative errors:
$\delta\widetilde{\Omega}_1/\widetilde{\Omega}_1$ (blue crosses),
$\delta\widetilde{\Omega}_2/\widetilde{\Omega}_2$ (solid-red line)
and $\delta T/T$ (dashed-green line).}
\end{figure}
Seen from the dashed-green line in Fig. 4, we find that the scheme
is quite robust against the timing errors, i.e., when $|\delta
T/T|\leq10\%$, we have $P_3(T')\geq0.9956$. Besides, according to
the blue crosses and the solid-red line in Fig. 4, the influences of
pulses' errors are larger than the timing error, however, $P_3(T)$
keeps higher than 0.9745 when
$|\delta\widetilde{\Omega}_1/\widetilde{\Omega}_1|\leq10\%$ or
$|\delta\widetilde{\Omega}_2/\widetilde{\Omega}_2|\leq10\%$.
Therefore, the scheme holds nice robustness against the parameters'
errors.

Secondly, let us analyze the robustness of the scheme against the
decoherent factors. Here, we consider a superconducting (SC) qubit
with $\Lambda$-type structure. For the SC qubit, there exists four
decoherent factors: (i) the energy relaxation for the path
$|2\rangle\rightarrow|1\rangle$ with energy relaxation rate
$\Gamma_1$, (ii) the energy relaxation for the path
$|2\rangle\rightarrow|3\rangle$ with energy relaxation rate
$\Gamma_2$, (iii) the dephasing between energy levels $|2\rangle$
and $|1\rangle$ with dephasing rate $\Gamma_{\phi1}$,  (iv) the
dephasing between energy levels $|2\rangle$ and $|3\rangle$ with
dephasing rate $\Gamma_{\phi2}$. Therefore, the evolution of the SC
qubits can be described by a master equation in Lindblad form as
following
\begin{equation}\label{d1}
\dot{\rho}=i[\rho,H_0]+\sum\limits_{l}[L_l\rho
L_l^{\dagger}-\frac{1}{2}(L_l^{\dagger}L_l\rho+\rho
L_l^{\dagger}L_l)],
\end{equation}
where, $L_l$ ($l=1,2,3,4$) is the Lindblad operator. Here, we have
four Lindblad operators as
\begin{eqnarray}\label{d2}
&&L_{1}=\sqrt{\Gamma_1}|1\rangle\langle2|,\ \ \
L_{2}=\sqrt{\Gamma_2}|3\rangle\langle2|,\cr\cr&&
L_3=\sqrt{\Gamma_{\phi1}}(|2\rangle\langle2|-|1\rangle\langle1|),\ \
\ L_4=\sqrt{\Gamma_{\phi2}}(|2\rangle\langle2|-|3\rangle\langle3|).
\end{eqnarray}
We plot $P_3(T)$ versus $\Gamma_1/\widetilde{\Omega}_0$ and
$\Gamma_2/\widetilde{\Omega}_0$
($\Gamma_{\phi1}/\widetilde{\Omega}_0$ and
$\Gamma_{\phi2}/\widetilde{\Omega}_0$) in Fig. 5 (a) (Fig. 5 (b)).
\begin{figure}
\scalebox{0.8}{\includegraphics[scale=0.8]{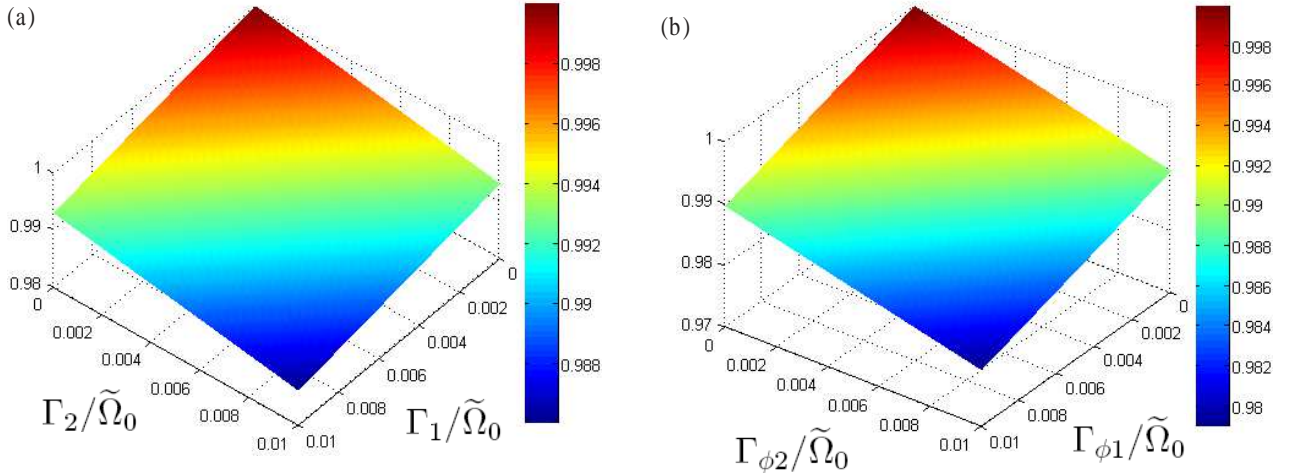}}
\caption{(a)$P_3(T)$ versus $\Gamma_1/\widetilde{\Omega}_0$ and
$\Gamma_2/\widetilde{\Omega}_0$. (b)$P_3(T)$ versus
$\Gamma_{\phi1}/\widetilde{\Omega}_0$ and
$\Gamma_{\phi2}/\widetilde{\Omega}_0$.}
\end{figure}
Seen from Fig. 5 (a), $P_3(T)$ keeps higher than 0.986 for all
$\Gamma_1$ and $\Gamma_2$ satisfying
$\Gamma_1/\widetilde{\Omega}_0\leq0.01$ and
$\Gamma_2/\widetilde{\Omega}_0\leq0.01$. According to Fig. 5 (b), we
have $P_3(T)\geq0.979$ when
$\Gamma_{\phi1}/\widetilde{\Omega}_0\leq0.01$ and
$\Gamma_{\phi2}/\widetilde{\Omega}_0\leq0.01$. Therefore, the scheme
is also quite robust against energy relaxations and dephasings for
SC qubits. However, the SC qubits is more sensitive to dephasings
when using STIRAP, with pulses shown in Eq.~(\ref{e18}), and
parameters $\Omega_0=45/T$, $t_0=0.15T$, $t_c=0.20T$, we have
$P_3(T)=0.9561$ when
$\Gamma_{\phi1}/\Omega_0=\Gamma_{\phi2}/\Omega_0=0.01$. According to
Refs. \cite{WuQIP,WeiQIP14}, for a multi-qubit system which has an
effective Hamiltonian in $\Lambda$-type structure, the dephasings
influence the SQ very much when using STIRAP. For example, Ref.
\cite{WuQIP} has shown that with STIRAP, when the ratio between
dephasing and coupling strength is only 0.0001, the fidelity of the
target state falls from 1 to about 0.85. Therefore, the scheme may
help to improve STIRAP for SC qubits.

\section{conclusion}

In conclusion, we have proposed an alternative scheme to construct a
shortcut to the adiabatic passage via picture transformation for
quantum transfer in a system with three-level $\Lambda$-type
structure. Different from previous works
\cite{BaksicPRL116,GaraotPRA89,OpatrnyNJP16,SaberiPRA90,TorronteguiPRA89,TorosovPRA87,TorosovPRA89,yehongPRA93,yehongarxiv,KangSR6,IbanezPRA87,IbanezPRL109,SongPRA93,HuangLPL13},
the present scheme has its own feature. For example, schemes
\cite{IbanezPRA87,IbanezPRL109,SongPRA93} have adopted a serials of
iterative picture transformations, while here picture transformation
is used only once. Besides, the ideas of schemes
\cite{IbanezPRA87,IbanezPRL109,SongPRA93} are to cancel the
transitions between the eigenstates of iterative Hamiltonian in each
iterative picture, and the idea of scheme \cite{BaksicPRL116}
suggests to cancel the transitions between a set of chosen dressed
states. But for the present scheme, we directly study the dynamics
of the three-level $\Lambda$-type system and the solution of the
Schr\"{o}dinger equation. Therefore, the present scheme consider the
way to construct STAP from a different viewpoint. The present scheme
has several advantages: (1) By choosing suitable control parameters,
experimentally feasible pulses can be designed. (2) The quantum
transfer can be achieved without any additional couplings. (3)
Comparing with quantum transfer with adiabatic passages, the
evolution is significantly sped up with present scheme.

Since the three-level $\Lambda$-type structure is very common in all
kinds of physical systems including superconducting qubits
\cite{ZhangSR5,WeiQIP14,WuQIP}, quantum dots or NV centers
\cite{SongNJP6,SongPRA93}, boson gas in longitudinal coordinate
coupled waveguides
\cite{GaraotPRA89,LonghiLPR3,LonghiJPB44,OrnigottiJPB41,RangelovPRA85,FisherPRB40,JakschPRL81},
atoms trapped in the cavities
\cite{yehongOC140,chenzhenSR6,lumeiPRA89,yehongPRA89,yehongSR5,xiaobinQIP14,wujiangQIP15,yehongPRA91},
etc., the present scheme can be a choice to construct STAP in these
physical systems. Considering the potential applications of the
present scheme in experiments, we hope the present scheme may be
useful in quantum information field.

\section*{ACKNOWLEDGEMENT}

This work was supported by the National Natural Science Foundation
of China under Grants No. 11575045, No. 11374054 and No. 11675046,
and the Major State Basic Research Development Program of China
under Grant No. 2012CB921601.

\end{document}